\documentclass[10pt]{iopart}
\usepackage{graphicx,epsfig}
\usepackage{bm}
\usepackage{iopams}
\usepackage{color,ulem}
\newcommand{\ba}{\begin{eqnarray}}
\newcommand{\ea}{\end{eqnarray}}

\newcommand{\dd}{{\textrm {d}}}

\usepackage{pdfsync}
\begin{document}



\title [Relativistic interpretation and cosmological signature of Milgrom's acceleration. ] {Relativistic interpretation and cosmological signature of Milgrom's acceleration.} 
  

\author{Roberto A. Sussman$^1$}
\address{Instituto de Ciencias Nucleares, Universidad Nacional Aut\'onoma de M\'exico (ICN-UNAM),
Apartado Postal 70--543, 04510 Ciudad de M\'exico, M\'exico.}

\author{X. Hernandez$^2$}
\address{Instituto de Astronom\'{\i}a, Universidad Nacional Aut\'{o}noma de M\'{e}xico (IA-UNAM),
Apartado Postal 70--264 C.P. 04510 Ciudad de M\'exico, M\'exico.}

\ead{$^1$sussman@nucleares.unam.mx, $^2$xavier@astro.unam.mx }




\date{Released 17 May 2017}

%





\begin{abstract}
  We propose in this letter a relativistic coordinate independent interpretation for Milgrom's acceleration
  $a_{0}=1.2 \times 10^{-8} \hbox{cm/s}^{2}$ through a geometric constraint obtained from the product of the
  Kretschmann invariant scalar times the surface area of 2--spheres defined through suitable characteristic length
  scales for local and cosmic regimes, described by Schwarzschild and Friedman--Lema\^\i tre--Robertson--Walker (FLRW) geometries, respectively. By demanding consistency between these regimes we obtain an
  appealing expression for the empirical (so far unexplained) relation between the accelerations $a_0$ and $c H_0$.
  Imposing this covariant geometric criterion upon a FLRW model, yields a dynamical equation for the Hubble
  scalar whose solution matches, to a very high accuracy, the cosmic expansion rate of
  the $\Lambda$CDM concordance model fit for cosmic times close to the present epoch. We believe that this geometric
  interpretation of $a_0$ could provide relevant information for a deeper understanding of gravity.

\end{abstract}

\maketitle


\section{Introduction} \label{intro}

The quantity $a_{0}=1.2 \times 10^{-8} \hbox{cm/s}^{2}$ has long been known in the astrophysical literature as Milgrom's
acceleration \cite{milg1983,milg1984,milg2002}. It has been used as a critical acceleration in the context of early
attempts to describe galactic dynamics without resorting to a dominant dark matter component, but rather in terms of a
change in gravitational physics becoming relevant at acceleration scales below $a_{0}$, all of which constitutes the basis for the  mostly empirical formulation, known as ``Modified Newtonian Dynamics'' (MOND), that fits the data on scales
beyond $R_M$ (see equation (\ref{scales})), that are characterised by accelerations below $a_0$ \cite{milg1984}. Indeed, the empirical scalings
of MOND, first calibrated from analysis of rotation curves of centrifugally supported spiral galaxies, have recently
been shown to apply also to the low acceleration regimes of very distinct classes of systems across over 10 orders of
magnitude in mass (elliptical galaxies, local globular clusters, dwarf galaxies in the Milky Way
and even wide binary star kinematics as measured by the Gaia satellite \cite{herna2012,scar2003,herna2017,herna2010,  lugha2014,jime2013,tian2016,macga1998,lelli2016,dabri2016,lelli2017,dura2017}).

Within the context of recent covariant extensions to GR constructed to reproduce the MOND phenomenology as a low velocity limit
\cite{Sergio, Salvatore}, the regime change from GR at high acceleration (and both low and high velocities) and the modified
covariant regime at low accelerations (and both high and low velocities, where in the latter MOND is recovered) is
introduced by hand, with a theoretical explanation for this transition still lacking. Further, the fact that $a_0$ is of the
same order of magnitude as $cH_0$ remains so far unexplained. This has motivated a more
recent approach to $a_0$ in a completely different context: the  ``emergent'' gravity theory proposed by  Verlinde
\cite{mccu2017,verl2016} in which the relation ``$a_0=cH_0$'', taken as an equality and denoted the ``Hubble acceleration'',
plays a central conceptual role within a novel theoretical approach, supported by ``insights'' from quantum information
theory, black hole physics and string theory. Other proposals to endow a theoretical interpretation for $a_0$ are found
in \cite{spec1,spec2}. However, all these proposals are still in their early stages and thus remain highly speculative. 

As an alternative theoretical approach to the ones summarized above, we present in this letter a proposal for a coordinate independent geometric interpretation of $a_0$ within the framework of metric gravity theories. For this purpose, we consider
relating this acceleration to a suitable geometric quantity related to the Kretschmann scalar,
$K\equiv {\cal R}_{\mu\nu\alpha\beta}{\cal R}^{\mu\nu\alpha\beta}$, which is the most fundamental curvature scalar that contains
the Ricci and Weyl contributions to curvature, and thus it should be nonzero in all non--trivial solutions of metric gravity
theories, including General Relativity (GR)
\footnote{The Kretschmann scalar appears in the quadratic curvature invariant ${\cal R}-4{\cal R}_{\mu\nu}{\cal R}^{\mu\nu} -K$ of Gauss--Bonnet gravity theory. However, in 4--dimensional manifolds the action from this invariant does not contribute to the dynamics because it becomes a total derivative \cite{GBinv}. }
.

Assuming $a_0$ and $c$ as fundamental constants of kinematic nature, dimensional analysis shows that the simplest quantity
with units of $\hbox{cm}^{-2}$ that can be formed by them is the ratio $a_0^2/c^4$. Since the Kretschmann scalar has units
$\hbox{cm}^{-4}$, the simplest geometric quantity based on this scalar with units $\hbox{cm}^{-2}$ follows by multiplying
it by a surface area and this product should then be matched to the constant ratio $a_0^2/c^4$. This suggests proposing
the conservation, along a physically motivated congruence of observers in self gravitating systems, of the product
\begin{equation}\kappa \equiv 4\pi\ell^2\times K(\ell),\label{kappadef}\end{equation}
where $\ell$ is suitable length scale characteristic of local or cosmic scales that should be described by appropriate
metrics to compute $K(\ell)$. Notice that  (\ref{kappadef}) is a purely geometric constraint that is independent on the
choice of a specific metric theory and/or any assumptions on the the matter--energy sources enclosed by 2--spheres of
surface area $4\pi\ell^2$. As we show along this letter, (\ref{kappadef}) yields an expression for $a_0$ that is
independent of the mass of local sources and is also consistent with cosmic dynamics as tested by observable cosmological
parameters within a FLRW context. Further, (\ref{kappadef}) can be useful to develop new insights as a constraint on modified
gravity theories that could generalize empiric MOND constructions without assuming the existence of dark matter.         
   
It is important to mention that the constraint (\ref{kappadef}) is different from the ``Bounding Curvature Constraint'', which
we presented and discussed in a recent paper \cite{BCC2018} with the aim of providing for stationary galactic systems a
geometric interpretation for $a_0$ that is also consistent with MOND dynamics in scales beyond $R_M$.   

\section{Milgrom and Schwarzschild scales}     

\begin{figure}
\includegraphics[height=7cm,width=7cm]{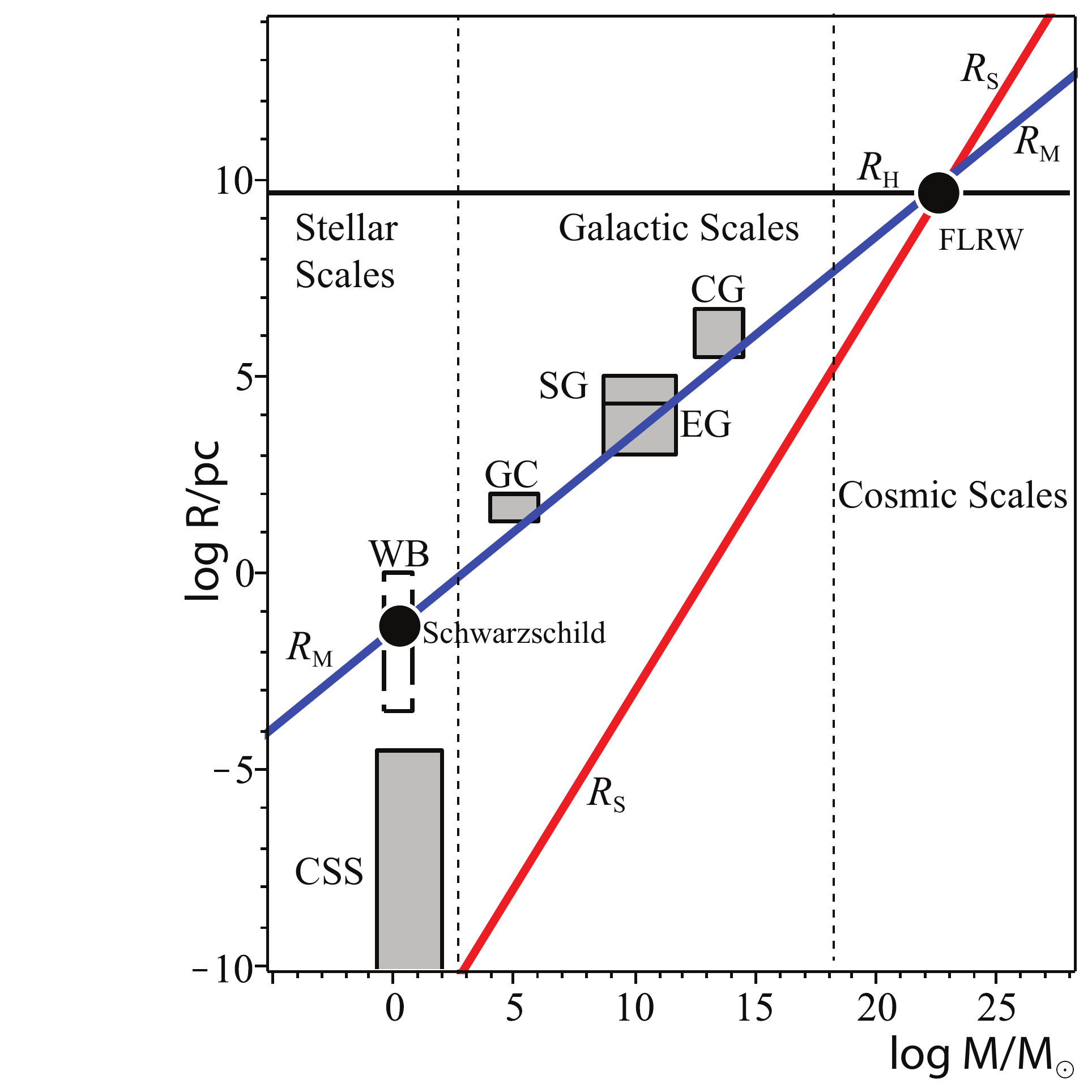}
\caption{Phase Space $R$ vs $M$ diagram. The red and blue lines respectively denote the Schwarzschild and Milgrom radii
  $R_S$ and $R_M$, the thick horizontal black line is the Hubble radius $R_H$ at $t=t_0$. The abbreviations CSS, WB, GC, EG, SG and GC respectively stand for compact stellar systems, wide binaries, globular clusters, elliptic galaxies, spiral galaxies and clusters of galaxies. Masses and characteristic radii correspond only to visible matter (see references cited in \cite{phase}). }
\label{fig1}
\end{figure}

In order to select an appropriate metric to evaluate (\ref{kappadef}) it is useful to examine the relation between Mass vs Radius
for various self--gravitating systems displayed in figure \ref{fig1} involving the two scales: $R_M$ and the Schwarzschild radius
$R_S$, first presented in \cite{phase}.
\begin{equation} R_M = \left(\frac{ GM}{a_{0}} \right)^{1/2},\qquad R_S = \frac{2GM}{c^2}.\label{scales}\end{equation}  
together with the present cosmic time Hubble radius $R_{H_0}$. As shown
in figure \ref{fig1}, giving the range of total baryonic masses extents and characateristic length scales $\ell$ for the various classes of systems shown,
the scales (\ref{scales}) arrange these systems along the following patterns: stellar scales (surroundings of isolated stars and compact multiple star systems) 
characterised by masses up to $M\sim 100\,\hbox{M}_\odot$ that are fully enclosed within $R_M$ ({\it i.e.} $R_S\ll \ell\ll R_M$);
cosmic scales (around the Hubble horizon) in which we can identify the characteristic
scale $\ell=\ell_0$ at present cosmic time in (\ref{kappadef}) with $R_{H_0}$, as it complies with $R_M=R_S \approx R_{H_0}$,  and intermediate galactic scales
with $R_s\ll \ell\sim R_M\ll R_{H_0}$. The three lines giving the mass dependences of $R_S$ and $R_M$, and the current value of
$R_{H_0}$, which (to current observationally accuracy) intersect at cosmic scales described by an FLRW metric
with $R_M\approx R_{H_0}$. All galactic systems extend beyond $R_M$, whereas compact stellar systems (CSS) are entirely contained within
$R_M$ and this scale is located in their weak Schwarzschild field. Wide extended binaries (WB) are a special case that we discuss further ahead.  

Evidently, probing (\ref{kappadef}) in a way that incorporates $R_M$ is easier in stellar and cosmic scales (see circles
in figure \ref{fig1}) that allow for a good approximation of the dynamics through simplified and idealized spacetime
metrics: solutions of either GR or any other metric theory. For compact stellar systems this suggests
computing $K(\ell)$ for a weak field Schwarzschild metric (black circle at the left in figure \ref{fig1}) and evaluating the product at $\ell\propto R_M$, since for
such systems $R_M\gg \ell$, so that we can ignore at scales around $R_M$  all structural details and describe these
systems as point sources in such field (rectangle marked by CSS in figure \ref{fig1}).  Likewise, we can follow the same
steps for probing (\ref{kappadef}) at cosmic scales of the order of magnitude of the Hubble horizon: compute (\ref{kappadef})
for an FLRW spacetime metric (thick black dot at the right in figure \ref{fig1}) and evaluate the product also at $R_M$ and $t=t_0$, which considering that $\ell_0=R_{H_0}\approx R_M$, means evaluating this product at the Hubble horizon (intersection of three length
scales in figure \ref{fig1}). 

The same procedure described above to probe (\ref{kappadef}) can be undertaken for any viable alternative gravity theory
by using its solutions (metrics) that describe  far fields of compact stellar systems and cosmic scales, as the latter must fit the same observations at these scales that have been successfully fit by GR solutions given by the Schwarzschild weak field
and FLRW metrics. In other words, solutions of the field equation of any viable alternative metric theory should be
quantitatively close to those of GR for compact stellar systems and cosmic scales, in the latter case, once the dynamically dominant
dark energy and dark matter hypothetical components are calibrated so as to match astronomical observations.    

For intermediate galactic scales, probing (\ref{kappadef}) becomes a more complicated task, as in this case it is
harder to select an exact appropriate spacetime metric to compute $K(\ell)$ in the region where $R_M$ is located in order to
evaluate this constraint. The reason is that (as shown by figure \ref{fig1}) $R_M$ lies within the self--gravitating body
in the midst of a matter--energy distribution that is much harder to describe through simple metrics.  

It is important to mention that $\ell\ll R_M$ does not hold for wide extended binaries \cite{herna2012,pittordis,herna19}. In this case we have two stars separated from each other by very large distances comparable or larger than the values $R_M$ for each of the individual stars.  Hence, the characteristic scale $\ell$ must be associated with the average distance
of the individual stars to the center of mass (see rectangle marked by WB in figure \ref{fig1}). Since we do have an effective 2--body system at $\ell =R_M$, we cannot use the weak field Schwarzschild metric to compute $K(\ell)$ as in compact stellar systems. This situation is more similar to that of galactic systems. Moreover, recent research \cite{herna19} shows changes in the dynamics of extended binaries at characteristic
scales $R_M$, thus suggesting that analogous effects could also occur in isolated stellar systems at these scales and beyond,
though this remains speculative as there is currently no observational evidence of this happening.

\section{Milgrom's acceleration for local isolated sources}        

These systems include isolated stars (from neutron stars to red giants) with typical radii of $\ell\sim 10-10^9$ km and masses in the range $M\sim 0.01-100\,\hbox{M}_\odot$, as well as compact binaries for which $\ell$ can be qualitatively similar to the average center of mass distance of individual stars
(of the order of several hundreds astronomical units). For such systems Milgrom's length scale $R_M$ is of the order of 0.03 pc, in the far field well beyond $\ell$.       

A first order description of the outer weak field of isolated stellar sources is furnished by the Schwarzschild weak field metric:
\begin{eqnarray}\fl
ds^{2} = -\left(1-\frac{2 G M}{c^{2} r}\right)\,c^{2}dt^{2} 
+\left(1+\frac{2 G M}{c^{2} r} \right)\,dr^{2} +r^{2}\left( d\theta^{2}+\sin^{2}\theta d\varphi^{2} \right),\label{Schw} 
\end{eqnarray}
where we have assumed that $r\gg 2GM/c^2$ holds and is also much larger than the characteristic scale $\ell$. To probe the
geometric constraint (\ref{kappadef}) we consider the family of 2--spheres, parametrized by the curvature radius $r$, generated by the intersection of the rest frames of static observers in (\ref{Schw}) (moving along a timelike Killing field) and the world tube in 4--dimensional spacetime generated by the worldlines
of these observers. The product of the Kretschmann curvature scalar
for  (\ref{Schw}) times the surface area of the 2--spheres is:
\begin{equation}
4\pi r^2 K=\frac{192\pi G^2M^2}{c^4 r^4},
\end{equation}
and should be evaluated at the characteristic  radius $r=\ell = \alpha R_{M}$, where $\alpha>0$ is a proportionality constant of $O(1)$ that
we evaluate further ahead. The result is:
\begin{equation}
 \kappa =  \frac{192 \pi}{\alpha^4} \left( \frac{a_{0}}{c^{2}} \right)^2,\label{kappaS}
\end{equation}
which does provide an appealing coordinate independent geometric definition for $a_0$, as it holds universally for all
masses $M$ that are wholly contained within within $R_M$ in a weak Schwarzschild field (\ref{Schw}), irrespective of any
assumption on the type of matter making up the source.

\section{The cosmological context}

\subsection{Consistency between local and cosmic scales}

We  now apply the geometric constraint (\ref{kappadef}) to a cosmological context described by the metric of homogeneous
and isotropic spatially flat FLRW models
\footnote{If we assume nonzero spatial curvature as restricted by observational constraints $|\Omega_0^k|\leq O(10^{-3})\ll 1$ we obtain practically indistinguishable results from those of the spatially flat case examined here.}
. Thus, the metric is now given by:

\begin{equation}
ds^{2}=-c^{2}dt^{2}+ a^2(t) \left[dr^{2}+r^{2}\left(d\theta^{2}+\sin^{2}\theta d\varphi^{2}\right)\right],\label{flrw}
\end{equation}
where $H_0=[\dot a/a]_0$ has units of $\hbox{sec}^{-1}$ (a dot
and subindex ${}_0$ will denote time derivative and evaluation at present cosmic time, respectively, where $a(t_0)=1$). In order to probe (\ref{kappadef}), we need to compute the Kretschmann scalar for the FLRW metric (\ref{flrw}) and multiply it by the surface area of a suitable collection of
2--spheres associated with a characteristic FLRW length scale. To keep a consistent approach to that followed for local
sources, we should evaluate (\ref{kappadef}) at $\ell=R_M$, but as shown in figure \ref{fig1} at present day cosmic scales we have
$R_M\approx R_{H_0}$, which suggests using the time dependent Hubble radius $R_H$:   
\begin{equation} \ell = R_H=\frac{c}{H},\qquad  H=\frac{\dot a}{a} = \frac13 \nabla_a u^a,\end{equation}
which is the most fundamental length scale for an FLRW metric, as it can be defined in a covariant manner as the divergence of the
4--velocity field $u^a$ of fundamental cosmic observers, and is independent of spatial curvature or assumptions on matter--
energy sources (independent even of the assumed metric gravity theory). The constraint (\ref{kappadef}) for 2--spheres
associated with $R_H$ becomes then 
\begin{equation}
\kappa=\frac{48\pi\,H^2}{c^2} \left(q^2+ 1 \right),\label{kappadef2}
\end{equation}
where the deceleration parameter $q$ is defined as $q\equiv -a\,\ddot a/\dot a^2=-(1+\dot H/H^2)$. To keep consistency with
the approach followed with local sources that resulted in (\ref{kappaS}), we demand that  (\ref{kappadef2}) be constant.
Hence, we impose the following conservation law preserving the constraint (\ref{kappadef}) now through the fluid flow
associated with fundamental cosmic observers, leading to the following very appealing form
\ba \fl  u^a\nabla_a\kappa=\dot\kappa=0 \quad\Rightarrow\quad \kappa = \kappa_0\quad\Rightarrow\quad H^2(q^2+1)=H_0^2(q_0^2+1).\label{kappaFflat}\ea
To be able to use this constraint we consider $q_0=-0.5275$, which  emerges from the Planck 2015 results \cite{planck} under the assumption of a fit to a $\Lambda$CDM model with matter (CDM plus baryons) density parameter $\Omega_0^{\textrm{\tiny{m}}}=0.315$ and $\Omega_0^\Lambda=0.685$. However, notice that we use this value for the sole purpose of calibration, as it is an empiric result obtained by observations that should be valid in any viable gravity theory under consideration. 

Comparing (\ref{kappaS}) with (\ref{kappaFflat}) we obtain the following expression relating $a_0$ with observable cosmological parameters
\begin{equation} a_0 = \frac{\alpha^2\sqrt{1+q_0^2}}{4\sqrt{3}}\times c\,H_0 \approx \frac{cH_0}{5.83},\label{a0RMforms}\end{equation}
where the third quantity is the well known numerical correspondence  between $a_0$ and $H_0\approx H_{70}$ \cite{milg2002}.
Considering the numerical value $1+q_0^2=1.277$ that we have used to calibrate the solutions emerging from the constraint
(\ref{kappaFflat}) through the latest observational data, we obtain for the proportionality constant $\alpha$ in
(\ref{a0RMforms}) the appealing value: $\alpha= 1.0511\approx 1$, that accounts for inaccuracies in the determination of
cosmological parameters and for the empiric numerical factor $1/5.83$. We have then     
\begin{equation} a_0 = \frac{\sqrt{1+q_0^2}}{4\sqrt{3}}\,cH_0,\quad R_M = \frac{2\times 3^{1/4}}{\left(1+q_0^2\right)^{1/4}}
  \left(\frac{GM}{cH_0}\right)^{1/2},\label{cH0}\end{equation}
which remarkably provides appealing theoretical forms for $a_0$ and $R_M$, quantities that have been hitherto understood only
in terms of empiric fitting formulae of Newtonian MOND. Notice that (\ref{cH0}) could also reveal an interesting potentially Machian effect in
which present day cosmic scale parameters $H_0,\,q_0$ imprint a signature on the dynamics of self--gravitational systems at
galactic scales (similar to the effects appearing in \cite{mannheim}).

\subsubsection{Fit to a $\Lambda$CDM model}

\begin{figure}
\includegraphics[height=3.5cm,width=7.5cm]{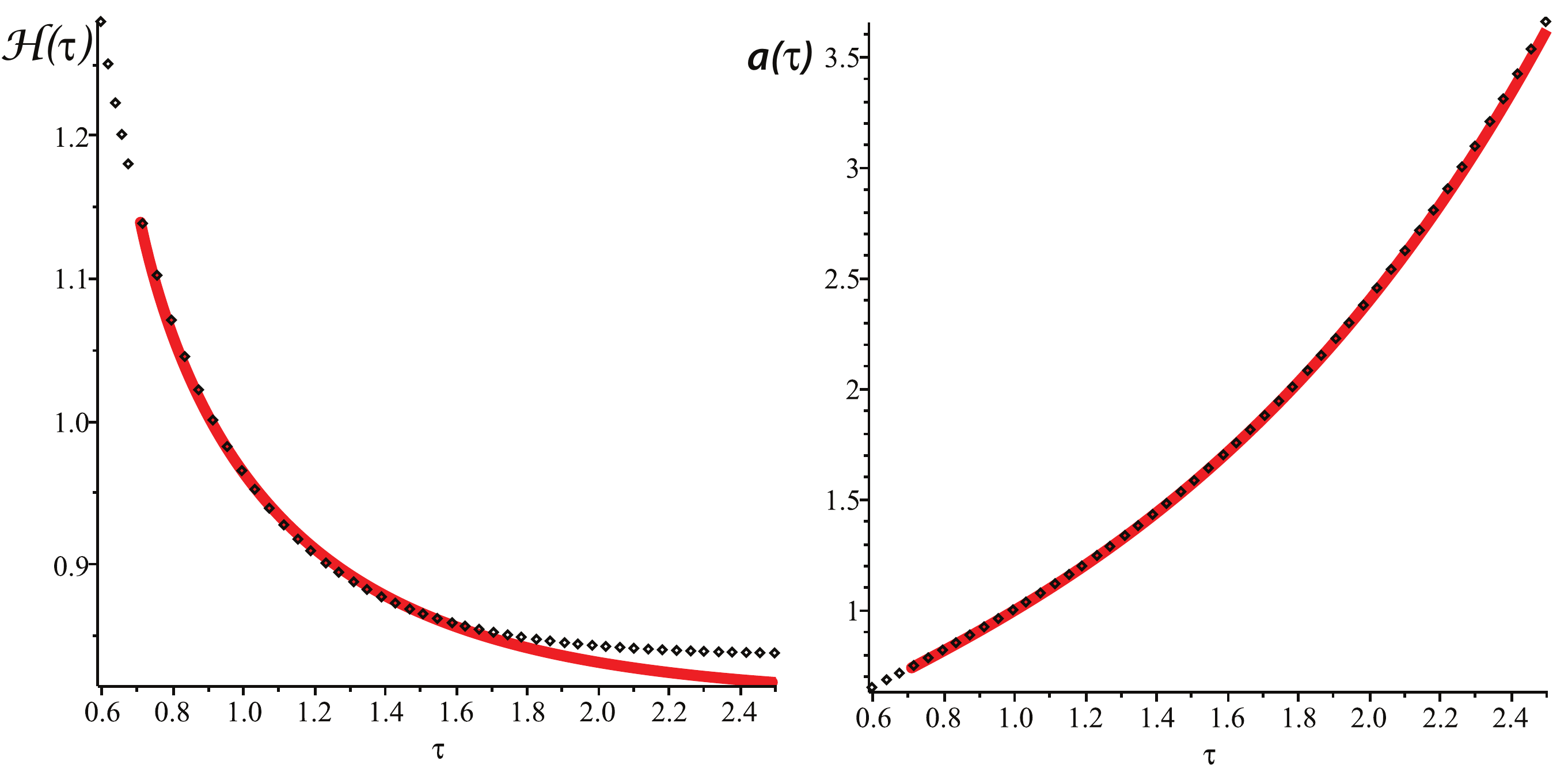}
\caption{Comparison between the expansion rate $\hat H(\hat t)$ (left panel) and scale factor $a(\hat t)$ (right panel), as
  predicted by the constraint (\ref{kappaFflat})(thick red curve) and as obtained from a $\Lambda$CDM model (dotted black curve).}
\label{fig2}
\end{figure}

\begin{figure}
\includegraphics[width=9.0cm,height=5.0cm]{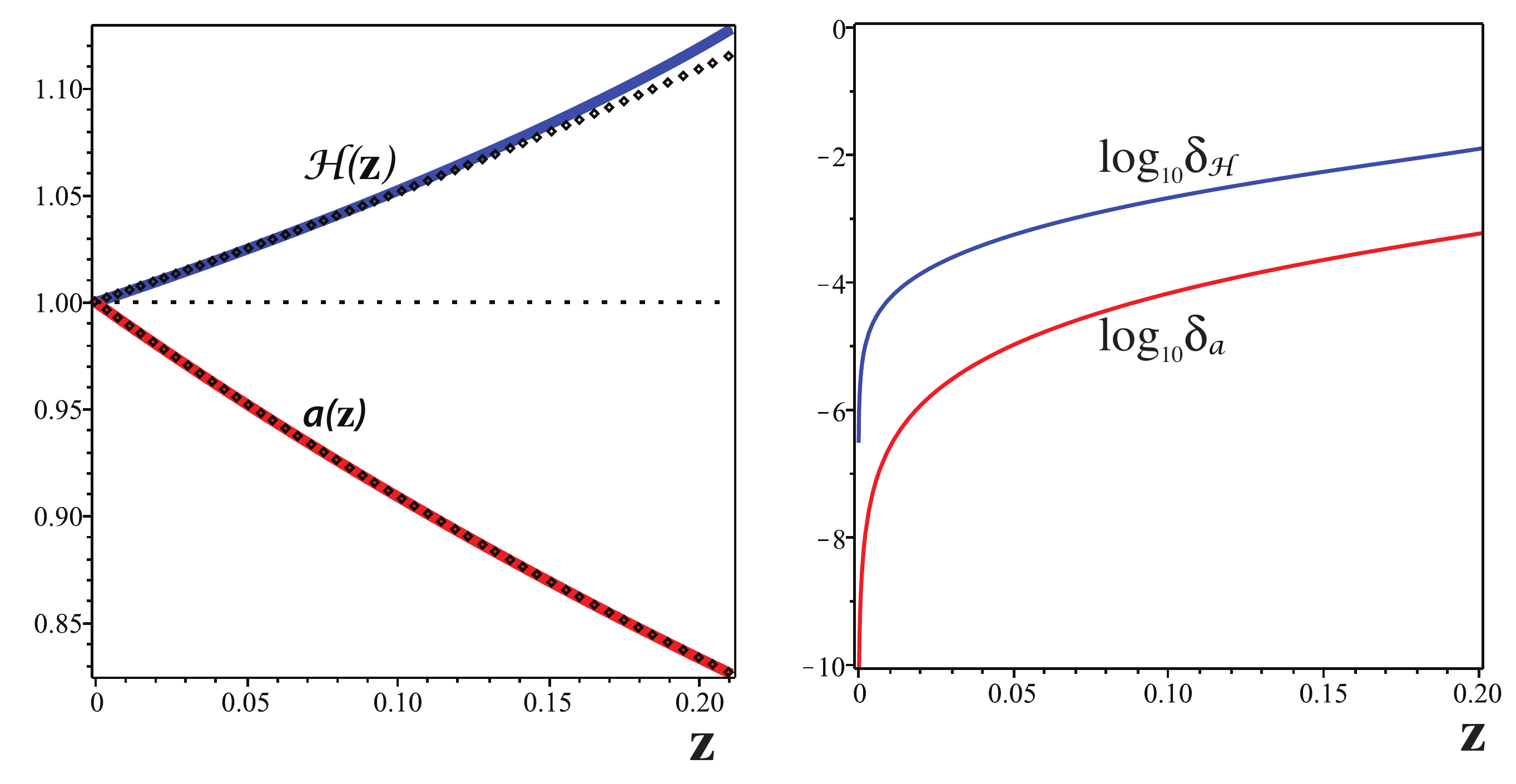}
\caption{The left panel displays the Hubble scalar $\hat H=H/H_0$ (curve in blue) and scale factor $a$ (curve in red) as
  functions of redshift from $z=0.2$, compare with their $\Lambda$CDM curve for parameters consistent with Planck 2015 values.
  Notice how the fit is much tighter for $a$ than for $\hat H$. The right panel displays  $\log_{10}\delta_{\hat H}$ and
  $\log_{10}\delta_a$, where $\delta_{\hat H}=|\hat H_{\textrm{\tiny{bcc}}}/\hat{H}-1|$ and $\delta_a=|a_{\textrm{\tiny{bcc}}}/a-1|$,
  with $\hat H_{\textrm{\tiny{bcc}}}(z),\,a_{\textrm{\tiny{bcc}}}(z)$ are the quantities obtained from (\ref{kappaFflat}) and
  $\hat H(z),\,a(z)$ their $\Lambda$CDM counterparts. Notice that $\delta_a\sim O(10^{-5})$ and $\delta_{\hat H}\sim O(10^{-3})$,
  which denotes a very tight fit to the $\Lambda$CDM curves. }
\label{fig3}
\end{figure}

We explore now the possible connections between (\ref{kappaFflat}) and cosmic dynamics. Rewriting this equation in terms of $\dot H$ and $\dot a=aH$ and introducing the dimensionless parameters $\hat t =H_0t$ and $\hat H=H/H_0$ leads to the following differential equation
\ba \frac{\dd \hat H}{\dd\hat t}&=&-\hat H^2\pm\sqrt{\gamma_0-\hat H^2}\,\hat H,\quad \gamma_0=1+q_0^2=1.278,\label{eq21}\ea
that can be solved numerically for observed values of $q_0$ and initial conditions $\hat H_0=a(\hat t_0)=1$ for $t_0=13.7$
Gyr. Since $\dd\hat H/\dd\hat t+\hat H^2=\dd^2 a/\dd\hat t^2$, it is necessary to select the ``+'' sign in the square root
in (\ref{eq21}) to obtain a late time accelerated expansion. We examine below the predicted expansion rate $\hat H(\hat t\,)$ and the scale
factor $a(\hat t\,)$ obtained from solving numerically (\ref{eq21}).  
These functions must be compared to their equivalents in a viable gravity theory. Since any proposed gravity theory should reproduce, for times close to $\hat t_0=H_0t_0$, a cosmic evolution close to that of the  $\Lambda$CDM model of GR, we compare (for calibration purposes) the solutions of (\ref{eq21}) with those of the $\Lambda$CDM Raychaudhuri equation
\begin{equation}\frac{\dd \hat H}{\dd\hat t}=-\hat H^2-\frac{\Omega_0^{\textrm{\tiny{m}}}}{2a^3}+\Omega_0^\Lambda,\label{raych}\end{equation}
for the parameters $\Omega_0^{\textrm{\tiny{m}}}=0.315,\,\,\Omega_0^\Lambda=0.685$. As shown in figure \ref{fig2}, the solutions of (\ref{eq21}) predict forms for $\hat H(\hat t\,)$ and $a(\hat t\,)$ that closely match those
obtained for $\Lambda$CDM solutions of (\ref{raych}) in their late time evolution ($0.77 <\hat t<1.6$),
{\it i.e.} from the onset of the accelerated expansion. A more accurate description of the fit is displayed by plotting $\hat H(z)$ and
$a(z)$ in the close past range range $0<z<0.2$ (left panel of figure \ref{fig3}), while the right panel displays the logarithm of the
relative differences. Notice how the fit for $a$ is much tighter than that of $\hat H$, though the error in the Hubble factor
is still well under 1\% in this range of redshifts.

\subsubsection{Equation of state.}
An interesting comparison between the predictions of the $\Lambda$CDM model and those from the cosmological implications of
the constraint (\ref{kappaFflat}) comes from calculating the equation of state parameter, $w=p/\rho$, that
would result from an effective GR solution for which $H$ is given by solutions of (\ref{eq21}) that
emerges from this constraint. For a spatially flat FLRW source made up of dust--like matter (baryons plus CDM)
($p_{\textrm{\tiny{m}}}=0$) and a dark energy fluid satisfying $p_{\textrm{\tiny{de}}}=w\rho_{\textrm{\tiny{de}}}$ with $w=w(a)$, the
dimensionless Omega parameters satisfy: $\Omega^{\textrm{\tiny{m}}}(a) +\Omega^{\textrm{\tiny{de}}}(a)=1$. This leads to the 
Raychaudhuri equation $\hat H^{-2}\dd/\dd\hat t(\hat H)=-1-\frac12[1+3w(1-\Omega^m)]$, which we compare with (\ref{eq21}), leading to the following  link between $w,\,\hat H$ and $\Omega^{\textrm{\tiny{m}}}$:
\begin{equation} w =-\frac{2\hat H\sqrt{1+q^2-\hat H^2}+1}{3(1-\Omega^{\textrm{\tiny{m}}})}\quad\Rightarrow\quad w_0 =
  -\frac{2|q_0|+1}{3(1-\Omega_{0}^{\textrm{\tiny{m}}})},\label{eqw}\end{equation} 
where we used (\ref{kappaFflat}) and $\hat H_0=1$. Choosing (for calibration) the Planck value $\Omega_{0}^{\textrm{\tiny{m}}} = 0.315$ together with $q_0=-0.5275$, we obtain  $w_0=-1.0260$, which
is a very close fit to the $\Lambda$CDM value $w=w_0=-1$.  In fact, the slight deviation from $w_0=-1$ fits very well the various attempts
to estimate empirically a dynamical dark energy distinct from a cosmological constant \cite{varw}.  
%

\section{Concluding remarks}

We have provided an elegant geometric interpretation for Milgrom's acceleration $a_0$ by means of the preservation of $\kappa$,
defined in (\ref{kappadef}) as the product of the Kretschmann scalar invariant times the surface area of a collection of
2--spheres defined by physically motivated congruences of observers, whose radius is the characteristic length scale $R_H$
associated with $a_0$. This is a
purely geometric covariant constraint that does not depend on any assumption on the nature of matter sources, and can be,
in principle applicable to any self--gravitating system and in any metric gravity theory that fulfills the equivalence principle.  

We considered Schwarzschild and FLRW geometries to calibrate and probe constraint (\ref{kappadef}), selecting as
characteristic length scales the radius $R_M$ (Schwarzschild weak field of stellar systems) and the Hubble radius
$R_H\approx R_M$ that can be defined for any cosmological FLRW model. While the Schwarzschild and FLRW metrics are
GR solutions, they provide a very precise fit to observational data at solar system and cosmic scales. Thus, any proposed
alternative gravity theory must also be calibrated to fit this data at these same scales where these GR solutions yield
accurate descriptions (once dark matter and dark energy are assumed and calibrated at galactic and cosmic scales). 

By comparing $\kappa$ for small and large scales (calibrated by Schwarzschild and FLRW geometries), we obtained in (\ref{cH0})
a very appealing theoretical interpretation for the (still unexplained) empiric relation between the accelerations $a_0$ and
$cH_0$ and for the 'Newton to MOND' transition scale. This interpretation might provide a signature of observable cosmic scale
parameters $q_0,\,H_0$ in the dynamics of local systems in scales below $R_M$. The implications of (\ref{kappadef}) in cosmic
dynamics yields the expansion rate $H$ and scale factor $a$ of an FLRW model that closely mimmic those of the $\Lambda$CDM
model for cosmic times from the onset of the accelerated expansion (see figure \ref{fig2}). This fit is very accurate for times
and redshifts close to the present epoch (see figure \ref{fig3}).         

We fully acknowledge the limitations our our results: we have only tested this interpretation for Milgrom's acceleration in
the very basic and highly symmetric Schwarzschild and FLRW spacetimes. Still, this geometric constraint
could provide a useful insight for testing and constructing modified theories of gravity. Further, testing this proposal in galactic
scales remains an urgent unfinished task that requires further work: as shown in \cite{BCC2018}, it might be necessary to
modify the constraint (\ref{kappadef}) to provide also a satisfactory fit to the more complicated dynamics of galactic
systems (this possible modification is still work in progress). We are also considering  possible  theoretical connections
with lattice structure models \cite{cellmods1,cellmods2}, cosmological holographic proposals \cite{coshor} and the
``emergent'' gravity proposal \cite{mccu2017,verl2016}.

We believe that we have provided sufficient elements to question the possibility that the results we have presented follow
from a mere coincidence. Rather, we believe that these results provide a  useful clue for a better understanding
of gravity.

\section*{Acknowledgements}

XH acknowledges support from DGAPA-UNAM PAPIIT IN-104517 and CONACyT, and RAS acknowledges
support from CONACYT 239639 and PAPIIT-DGAPA RR107015.


\section*{References}

\begin{thebibliography}{99}

\bibitem{milg1983} Milgrom, M. 1983, ApJ, 270, 365
\bibitem{milg1984} Milgrom, M. 1984, ApJ, 287, 571
\bibitem{milg2002} Milgrom M., 2002, New Astron. Rev., 46, 741

\bibitem{covMOND} Bekenstein J. D., 2004, Phys. Rev. D, 70, 083509; Capozziello, S., \& de Laurentis, M. 2011, Phys. Rep., 509, 16 7; Mendoza S., Bernal T., Hernandez X., Hidalgo J. C., Torres L. A., 2013, MNRAS, 433, 1802; Moffat, J. W., \& Toth, V. T. 2008, ApJ, 680, 1158; Zhao, H., \& Famaey, B. 2010, PhRvD, 81, 087304; Capozziello S., Cardone V.F., Troisi A., 2007, MNRAS, 375, 1423

\bibitem{dabri2016} Dabringhausen, J., Kroupa, P., Famaey, B., Fellhauer, M. 2016, MNRAS, 463, 186

\bibitem{dura2017} Durazo R., Hernandez X., Cervantes Sodi B., S\'anchez S. F., 2017, ApJ, 837, 179

\bibitem{herna2010} Hernandez X., Mendoza S., Suarez T., Bernal T., 2010, A\&A, 514, 101  
\bibitem{herna2012} Hernandez X., Jim\'enez M. A., Allen C., 2012, Eur. Phys. J. C, 72, 1884
\bibitem{herna2017} Hernandez X., Cortes R. A. M., Scarpa R., 2017, MNRAS, 464, 2930
\bibitem{jime2013} Jim\'enez, M. A., Garcia, G., Hernandez, X., Nasser, L. 2013, ApJ, 768, 142
\bibitem{lelli2016} Lelli F., McGaugh S. S., Schombert J. M., 2016, AJ, 152, 157
\bibitem{lelli2017} Lelli, F., McGaugh, S. S., Schombert, J. M., Pawlowski, M. S. 2017, ApJ, 836, 152  
\bibitem{lugha2014} L\"{u}ghausen F., Famaey B., Kroupa P., 2014, MNRAS, 441, 2497
\bibitem{macga1998} McGaugh S. S., de Blok W. J. G., 1998, ApJ, 499, 66
\bibitem{scar2003} Scarpa R., Marconi G., Gilmozzi R., 2003, A\&A, 405, L15
\bibitem{tian2016} Tian, Y., Ko, C.-M. 2016, MNRAS, 462, 1092

\bibitem{Sergio}  E. Barrientos and S. Mendoza, Phys. Rev. D 98 (2018) 084033.

\bibitem{Salvatore} Capozziello S., Jovanovic P., Borka Jovanovic V., Borka D., 2017, JCAP, 06, 044

\bibitem{verl2016} Verlinde, E., 2016, preprint arXiv:1611.02269

\bibitem{mccu2017} McCulloch M., 2017, Ap\&SS, 362, 57

\bibitem{spec1} Bernal, T., Capozziello, S., Hidalgo, J. C., Mendoza, S. Eur. Phys. J. C, 2011,71, 1794-1801.

\bibitem{spec2} Bernal, T., Capozziello, S., Cristofano, G., de Laurentis, M. Mod. Phys. Lett. A, 2011,26, 2677-2687.

\bibitem{GBinv} Mardones A. and Zanelli J., 1991, Class. Quantum Grav., 8, 1545 

\bibitem{BCC2018} Hern\'andez, X., Sussman, R.A., Nasser, L., 2019, MNRAS, 483, 147

\bibitem{phase}  Hern\'andez X., 2012, Entropy, 14, 848

\bibitem{pittordis} Pittordis C., Sutherland W., 2018, MNRAS, 480, 1778

\bibitem{herna19} Hernandez X., Cortes R. A. M., Allen C., Scarpa R., 2019, International Journal of Modern Physics D, 28, 1950101










\bibitem{planck} Planck Contribution A\&A 594, A13 (2016)

\bibitem{mannheim} Mannheim P. D., 2006, Prog. Part. Nuc. Phys.,56, 340

\bibitem{varw} de Felice A., Nesseris S., Tsujikawa S., 2012, JCAP, 05, 029; Jassal H. K., Bagla J. S., Padmanabhan T., 2005, MNRAS, 356, L11; Postnikov S., Dainotti M. G., Hernandez X., Capozziello S., 2014, ApJ, 783, 126


\bibitem{cellmods1} T. Clifton and P. G. Ferreira, Phys. Rev. D 80, 103503 (2009); Phys. Rev. D 84, 109902 (2011); T. Clifton and
P. G. Ferreira, JCAP 0910, 26 (2009); T. Clifton, K. Rosquist and R. Tavakol, Phys. Rev. D 86, 043506 (2012); J.-P. Bruneton and J. Larena, Class. Quant. Grav. 29, 155001 (2012); T. Clifton, Class. Quant. Grav. 28, 164011 (2011); A.A.A. Sanghai and T. Clifton, Phys. Rev. D 91, 103532 (2015) (erratum: Phys. Rev. D 93, 089903 (2016)); A.A.A. Sanghai and T. Clifton, Phys. Rev. D 94, 023505 (2016)

\bibitem{cellmods2}  E. Bentivegna and M. Korzynski, Class. Quant. Grav. 29, 165007 (2012); E. Bentivegna, Class. Quant. Grav. 31, 035004 (2014); E. Bentivegna and M. Korzynski, Class. Quant. Grav. 30, 235008 (2013);  C.-M. Yoo, H. Abe, Y. Takamori and K.-i. Nakao, Phys. Rev. D 86, 044027 (2012); C.-M. Yoo, H. Okawa and K.-i. Nakao, Phys. Rev. Lett. 111, 161102 (2013); C.-M. Yoo and H. Okawa, Phys. Rev. D 89, 123502 (2014).

\bibitem{coshor} D. Bak and S. J Rey, Class. Quant. Grav.17, L83 (2000); R. G. Cai and L. M. Cao, Phys. Rev. D75, 064008 (2007); 
R. G. Cai and S P Kim, JHEP, 0502 050(2005); V. Faraoni, Phys. Rev. D84, 024003 (2011); D. W. Tian and I. Booth, Phys. Rev. D 92, 024001 (2015)


  
\end{thebibliography}
\end{document}